\documentclass[3p,preprint,11pt]{elsarticle}
\journal{Signal Processing}
\usepackage{amssymb,amsmath,amsthm,bm,url}

\biboptions{sort&compress}

\usepackage{graphicx}
\graphicspath{{figures-pdf/}}

\newlength\imagewidth
\newlength\figwidth
\usepackage[normalem]{ulem}

\begin{document}

\begin{frontmatter}

%\title{Breaking an Image Scrambling Encryption Algorithm of Pixel Bit Based on Chaos Map}
%\title{Security analysis of a binary image permutation scheme based on Logistic map}

\title{Optimal quantitative cryptanalysis of permutation-only multimedia ciphers against plaintext attacks}
    
\author[cn-xtu,hk-polyu]{Chengqing Li\corref{corr}}
\ead{chengqingg@gmail.com}
\author[hk-polyu]{Kwok-Tung Lo}
\cortext[corr]{Corresponding author.}

\address[cn-xtu]{College of Information Engineering, Xiangtan University, Xiangtan 411105, Hunan, China}
\address[hk-polyu]{Department of Electronic and Information Engineering, The Hong Kong Polytechnic University, Hong Kong}

\begin{abstract}
Recently, an image scrambling encryption
algorithm of pixel bit based on chaos map was proposed. Considering
the algorithm as a typical binary image scrambling/permutation
algorithm exerting on plaintext of size $M\times (8N)$, this paper
proposes a novel optimal method to break it with some
known/chosen-plaintexts. The spatial complexity and computational
complexity of the attack are only $O(32\cdot MN)$ and $O(16\cdot n_0\cdot MN)$
respectively, where $n_0$ is the number of known/chosen-plaintexts used.
The method can be easily extended to break any permutation-only encryption scheme exerting on
plaintext of size $M\times N$ and with $L$ different levels of
values. The corresponding spatial complexity and computational
complexity are only $O(MN)$ and $O(\lceil\log_L(MN)\rceil \cdot MN)$ respectively.
In addition, some specific remarks on the performance of the image scrambling encryption
algorithm are presented.
\end{abstract}


\end{frontmatter}

\section{Introduction}

With rapid development of digital information technology, image data
is transmitted over all kinds of wired/wireless channels more and
more frequently. Consequently, security of image data becomes more
and more important. However, the traditional text encryption schemes
fail to protect image data efficiently due to the big differences
between image data and text, e.g. strong redundancy existing in
uncompressed image data and its bulky size. In addition, image
encryption schemes have some special requirements such as fast handling
speed and easy concatenation of different components of the whole
image processing system. Therefore, designing encryption schemes protecting
image data specially becomes an urgent task. Due to the subtle
similarities between cryptography and chaos, a great number of image
encryption schemes based on chaos or other nonlinear theories have been proposed in the past
decade
\cite{Chung-Chang:SCAN:PRL1998,Chen&Yen:RCES:JSA2003,YaobinMao:CSF2004,
Flores:EncryptLatticeChaos06,Kurian:StreamCipher:SP08,Tong:Encrypt:SP09,LiaoXF:Wave:SP10}.
Unfortunately, most of them have been found to be insecure to
different extents from the viewpoint of modern cryptography
\cite{Chang&Yu:SCANImageCryptanalysis:PRL2002,Li:AttackDSEA2004,Kaiwang:PLA2005,
David:AttackingChaos08,Li:AttackingMaoScheme2007,Li:AttackingRCES2008,
Rhouma:BreakLian:PLA08,Li:AttackingIVC2009,Alvarez:BreakingNCA:CNSNS09}.
For more discussion on chaos-based image encryption schemes,
please refer to
\cite{AlvarezLi:Rules:IJBC2006,Li:ChaosImageVideoEncryption:Handbook2004}.

In \cite{Ye:Scramble:PRL10}, an image permutation algorithm was
proposed by scrambling/permuting binary bit of every pixel with
pseudo-random number sequence generated by chaotic logistic map.
Essentially, it is a permutation-only algorithm exerting on a binary
image of size $M\times (8N)$. The present paper focuses on security
analysis of the algorithm and reports the following results: 1) an
optimal method is proposed to break the image permutation algorithm
under study with some known/chosen plaintexts; 2) the method is
extended to break any permutation-only encryption scheme exerting on
elements of any different levels of values; 3) some remarks on the
performance of the image permutation algorithm under study are
given.

The rest of this paper is organized as follows. Section~\ref{sec:encryptscheme} briefly introduces the image permutation
algorithm under study. Section~\ref{sec:cryptanalysis} proposes the optimal known/chosen-plaintext attack based on a binary tree
to break the algorithm and points out some remarks on the performance on it. Extension of the optimal attack to break any
permutation-only multimedia encryption schemes is discussed in Sec.~\ref{sec:AttackAnyPermuation}. The last section
concludes the paper.

\section{The image permutation algorithm under study}
\label{sec:encryptscheme}

The plaintext encrypted by the image permutation algorithm under
study is a gray-scale image of size $M\times N$
(height$\times$width), which can be denoted by an $M\times N$ matrix
in domain $\mathbb{Z}_{256}$, $\bm{I}=[I(i,j)]_{i=0, j=0}^{M-1,
N-1}$. The image $\bm{I}$ is further represented as an $M\times
(8N)$ binary matrix $\bm{B}=[B(i,l)]_{i=0, l=0}^{M-1, 8N-1}$, where
$I(i, j)=\sum_{k=0}^7 B(i, l)\cdot 2^k$, $l=8\cdot j+k$. The
corresponding cipher-image is $\bm{I}'=[I'(i,j)]_{i=0, j=0}^{M-1,
N-1}$, $I'(i, j)=\sum_{k=0}^7 B'(i, l)\cdot 2^k$, $l=8\cdot j+k$.
Then, the image permutation algorithm can be described as
follows\footnote{To make the presentation more concise and complete,
some notations in the original paper are modified, and some details
about the algorithm are also supplied or corrected.}.

\begin{itemize}
\item \textit{The secret key}: three positive integers $m$, $n$, and $T$, and
the initial condition $x_0\in (0,1)$ and control parameter $\mu
\in(3.569945672, 4)$ of the following chaotic Logistic map:
\begin{equation}
f(x)=\mu\cdot x\cdot(1-x).
\label{eq:Logistic}
\end{equation}

\item \textit{The initialization procedure}: 1) run the map Eq.~(\ref{eq:Logistic})
from $x_0$ to generate a chaotic sequence,
$\{x_k\}_{k=1}^{\max{\{(m+M),(n+8MN)\}}}$; 2) generate an vector
$\bm{T_M}$ of length $M$, where $\bm{S_M}(\bm{T_M}(i))$ is the
$(i+1)$-th largest element of $\bm{S_M}=\{x_{m+k}\}_{k=1}^{M}$,
$0\leq i\leq (M-1)$; 3) generate a matrix $\bm{T_N}$ of size
$M\times (8N)$, where, $\forall$ $i\in\{0,\cdots, M-1\}$,
$\bm{S_N}(\bm{T_N}(i, j))$ is the $(j+1)$-th largest element of
$\bm{S_N}=\{x_{n+(8N)i+k}\}_{k=1}^{8N}$ , $0\leq j\leq N-1$.

\item \textit{The encryption procedure}:

\begin{itemize}
\item \textit{Step 1 -- vertical permutation}: generate an intermediate matrix
$\bm{B}^*=[B^*(i,l)]_{i=0, l=0}^{M-1, 8N-1}$, where
\begin{equation}
B^*(i,:)=B(\bm{T_M}(i),:);
\label{eq:encrypt_ver}
\end{equation}

\item \textit{Step 2 -- horizontal permutation}: generate an intermediate matrix
$\bm{B}'=[B'(i,l)]_{i=0, l=0}^{M-1, 8N-1}$, where
\begin{equation}
B'(i, l)=B^*(i,\bm{T_N}(i, l)); \label{eq:encrypt_hor}
\end{equation}

\item \textit{Step 3 -- repetition}: reset the value of $x_0$ to the current state of Eq.~(\ref{eq:Logistic}), and repeat
the above operations from \textit{the initialization procedure} for $(T-1)$ times.
\end{itemize}

\item \textit{The decryption procedure} is similar to the encryption one except the following simple modifications:
1) the different rounds of encryption are exerted in a reverse order; 2) the order of \textit{Step 1} and \textit{Step 2}
in each round is reversed; 3) the left parts and right parts of Eq~(\ref{eq:encrypt_ver}) and Eq.~(\ref{eq:encrypt_hor})
are exchanged, respectively.
\end{itemize}

\section{Cryptanalysis}
\label{sec:cryptanalysis}

\subsection{Known-plaintext attack}
\label{sec:KnownPlaintextAttack}

The known-plaintext attack is an attack model for cryptanalysis
where the attacker has some samples of both the plaintext and the
corresponding ciphertext and make use of them to reveal secret
information, such as secret keys and/or its equivalent ones.

Apparently, the combination of multiple rounds of the operations in
\textit{the encryption procedure} can be represented by an
$M\times(8N)$ permutation matrix $\bm{W}=[w(i, l)]_{i=0, l=0}^{M,
8N}$, where $w(i, l)=(i', l')$ denotes the secret position of the
plain bit $B(i, j)$ in $\bm{B}'$. That is, the permutation
matrix $\bm{W}$ defines a bijective map on set $\mathbb{M}\times
\mathbb{N^+}$, where $\mathbb{M}=\{0, \cdots, M-1\}$ and
$\mathbb{N^+}=\{0, \cdots, 8N-1\}$. Once the attacker recovers the
permutation matrix $\bm{W}$ and then obtains its inverse
$\bm{W}^{-1}$, he can use it as an equivalent key to decrypt any
cipher-image encrypted with the same secret key.

Since that the secret permutation does not change the
values of the permuted elements, a general algorithm was proposed in
\cite{Li:AttackingPOMC2008} for obtaining the permutation matrix by
comparing the values of the elements of some plaintexts and the
corresponding ciphertexts. Based on the same mechanism, a novel
optimal method is proposed here to break the image permutation
algorithm under study.

Actually, the proposed method is the construction process of a binary
tree, where every node includes five components in order: a pointer
holding address of the left child node, $PT_L$; two sets containing
some entry positions of plain-image and cipher-image, respectively;
cardinality of one of the two sets; a pointer holding address of the
right child node, $PT_R$. Denote the two sets in the root node with
$\mathbb{B}$ and $\mathbb{B}'$, respectively, where
$\mathbb{B}=\mathbb{B}'=\mathbb{M}\times \mathbb{N^+}$. Obviously,
cardinality of $\mathbb{B}$, $|\mathbb{B}|=8MN$. Then, the binary
tree can be constructed as follows.

\begin{itemize}
\item $\forall\ (i, l)\in \mathbb{B}$, do the
following operations:

\begin{equation}
\begin{cases}
\mbox{add } (i, l) \mbox{ into } \mathbb{B}_1 & \mbox{if }  B(i, l)=1; \\
\mbox{add } (i, l) \mbox{ into } \mathbb{B}_0 & \mbox{if } B(i,
l)=0.
\end{cases}
\label{eq:divide0}
\end{equation}

\item $\forall\ (i, l)\in \mathbb{B}'$, do the
following operations:

\begin{equation*}
\begin{cases}
\mbox{add } (i, l) \mbox{ into } \mathbb{B}'_1 & \mbox{if }  B'(i, l)=1; \\
\mbox{add } (i, l) \mbox{ into } \mathbb{B}'_0 & \mbox{if } B'(i,
l)=0.
\end{cases}
%\label{eq:divide1}
\end{equation*}

\item Delete the elements in the two sets of root node and set the
fourth item of the node as zero.
\end{itemize}

Since the secret permutation does not change the values of the
permuted elements, the cardinalities of the two sets in the two
child nodes are always the same. Hence, only the cardinality of one set
is needed to record. The basic structure of the binary tree is shown
in Fig.~\ref{fig:structure}. With more pairs of known-images and the
corresponding cipher-images, the binary tree will be updated and
expanded iteratively with the following steps.
\begin{itemize}
\item Search for all nodes whose third item is greater than one, namely,
both the corresponding two sets have more than one elements;

\item Expand each found node with the similar operations shown above.
\end{itemize}

\setlength{\unitlength}{1.1mm}
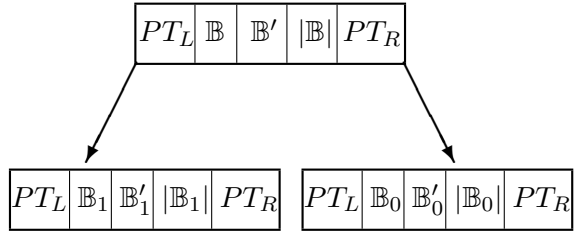
\begin{figure}[!htb]
\centering
\begin{picture}(72,27)
\thicklines\put(0,0){\framebox(32,7)[c]{$PT_L$ $\mathbb{B}_1$
$\mathbb{B}_1'$ $|\mathbb{B}_1|$ $PT_R$}}
\thinlines\put(7,0){\line(0,1){7}} \put(12,0){\line(0,1){7}}
\put(17,0){\line(0,1){7}} \put(24,0){\line(0,1){7}}

\thicklines\put(35,0){\framebox(32,7)[c]{$PT_L$ $\mathbb{B}_0$
$\mathbb{B}_0'$ $|\mathbb{B}_0|$ $PT_R$}}
\thinlines\put(42,0){\line(0,1){7}} \put(47,0){\line(0,1){7}}
\put(52,0){\line(0,1){7}} \put(59,0){\line(0,1){7}}

\thicklines\put(15,20){\framebox(32,7)[c]{$PT_L$ $\mathbb{B}$
$\mbox{\;}\mbox{\;}\mathbb{B}'$ $\mbox{\,}\mbox{\,}|\mathbb{B}|$ $
PT_R$}} \thinlines\put(22,20){\line(0,1){7}}
\put(27,20){\line(0,1){7}} \put(33,20){\line(0,1){7}}
\put(39,20){\line(0,1){7}}

\thicklines \put(15,20){\vector(-1,-2){6}}
\put(47,20){\vector(1,-2){6}}
\end{picture}
\caption{Basic structure of the binary tree.}
\label{fig:structure}
\end{figure}

After the construction of the binary tree is completed, we now investigate how
to obtain the estimated version of the permutation matrix $\bm{W}$
from the tree. To facilitate the following discussion, let
$\mathbb{B}_i$, $\mathbb{B}'_i$ and $|\mathbb{B}_i|$ denote the
middle three items of a leaf node, respectively. Apparently,
$\bm{w}(i, l)$ can be uniquely determined if and only if
$|\mathbb{B}_i|=1$. Otherwise, one has to guess one from
$|\mathbb{B}_i|!$ possible cases. For the whole permutation matrix,
there are $\Pi_{i=1}^{P}(|\mathbb{B}_i|!)$ possible cases, where $P$
is the number of leaf nodes in the binary tree. For simplicity, we
derive the permutation matrix by mapping the elements in
$\mathbb{B}_i$ and $\mathbb{B}'_i$ one by one in order.

Next, we investigate how many known-images are sufficient to achieve an
acceptable breaking performance. Roughly speaking,
$\Pi_{i=1}^{P}(|\mathbb{B}_i|!)$ rapidly decrease when the number of
known-images, $n_0$, is increased. The less value of
$\Pi_{i=1}^{P}(|\mathbb{B}_i|!)$ means more accurate estimation of
the permutation matrix. To simplify the analyses, we assume that
each element in $\bm{B}$ distributes uniformly over $\{0, 1\}$, and
any two elements are independent of each other. Then, the elements
of $\mathbb{B}_i$ can be divided into the following two types:
\begin{itemize}
\item \textit{the sole right position}, which occurs definitely;

\item \textit{other fake positions}, each of which occurs in $\mathbb{B}_i$
with a probability of $1/2^{n_0}$, since one condition in
Eq.~(\ref{eq:divide0}) is satisfied consecutively for $n_0$ times.
\end{itemize}

Let $n_8$ denote the number of error bits in a recovered pixel, it
can be calculated that
$Prob(n_8=i)=\binom{8}{i}\cdot(1-p_b)^i\cdot(p_b)^{8-i}$, where
$p_b=\frac{1}{1+(8MN-1)/2^{n_0}}$, $i=0\sim 8$. Generally speaking,
when every bit is determined correctly with a probability larger
than a half, namely $p_b>0.5$, the decryption will be acceptable.
Solve the inequality, one has
\begin{equation}
n_0>\lceil\log_2(8MN-1)\rceil. \label{eq:numberofplaintext}
\end{equation}

\begin{figure}[!htb]
\centering
\begin{minipage}[t]{\figwidth}
\centering
\includegraphics[width=\figwidth]{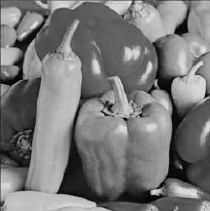}
a)
\end{minipage}
\begin{minipage}[t]{\figwidth}
\centering
\includegraphics[width=\figwidth]{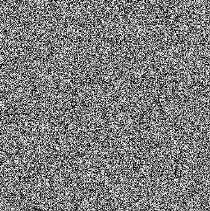}
b)
\end{minipage}\\
\begin{minipage}[t]{\figwidth}
\centering
\includegraphics[width=\figwidth]{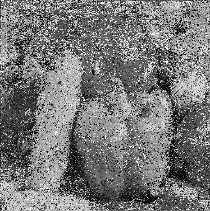}
c)
\end{minipage}
\begin{minipage}[t]{\figwidth}
\centering
\includegraphics[width=\figwidth]{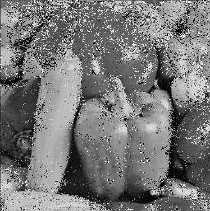}
d)
\end{minipage}\\
\caption{The image ``Peppers" recovered by know-plaintext attack: a)
``Peppers"; b) the encrypted ``Peppers"; c) recovered ``Peppers" via
$\bm{W}_{20}$; d) recovered ``Peppers" via $\bm{W}_{25}$.}
\label{figure:KnownPlainTextAttack}
\end{figure}

To verify the performance of the above known-plaintext attack, a
number of experiments have been performed on a number of
randomly selected natural images of size $256\times 256$. In this
case, Eq.~(\ref{eq:numberofplaintext}) become
$n_0>\lceil\log_2(2^{19}-1)\rceil=19$. With a randomly selected secret
key $(x_0, \mu, m, n, T)=(0.2009, 3.98, 20, 51, 4)$, a plain-image
``Peppers" and its encrypted version are shown in
Figs.~\ref{figure:KnownPlainTextAttack}a) and b), respectively. Let
$\bm{W}_{20}$ and $\bm{W}_{25}$ denote the estimated version of the
permutation matrix $\bm{W}$ obtained from 20 and 25 known
plain-images, respectively. The decrypted version of
Fig.~\ref{figure:KnownPlainTextAttack}b) with $\bm{W}_{20}$ and
$\bm{W}_{25}$ is shown in Figs.~\ref{figure:KnownPlainTextAttack}c)
and d) respectively. It is found that most visual information
contained in the original plain-image has been recovered in
Fig.~\ref{figure:KnownPlainTextAttack}c), although only
$23509/65536\approx 35.8\%$ of plain pixels are correct in value.
This is attributed to the following two main reasons:

\begin{itemize}
\item human eyes have a powerful capability for excluding image noises
and recognizing significant information;

\item due to strong redundancy in natural images, two pixel values
are close to each other with a probability larger than the average
one, hence many incorrectly recovered pixels are close to their true
values with a probability larger than the average one also.
\end{itemize}

\begin{figure}
\centering
\begin{minipage}[t]{1.5\imagewidth}
\centering
\includegraphics[width=1.5\imagewidth]{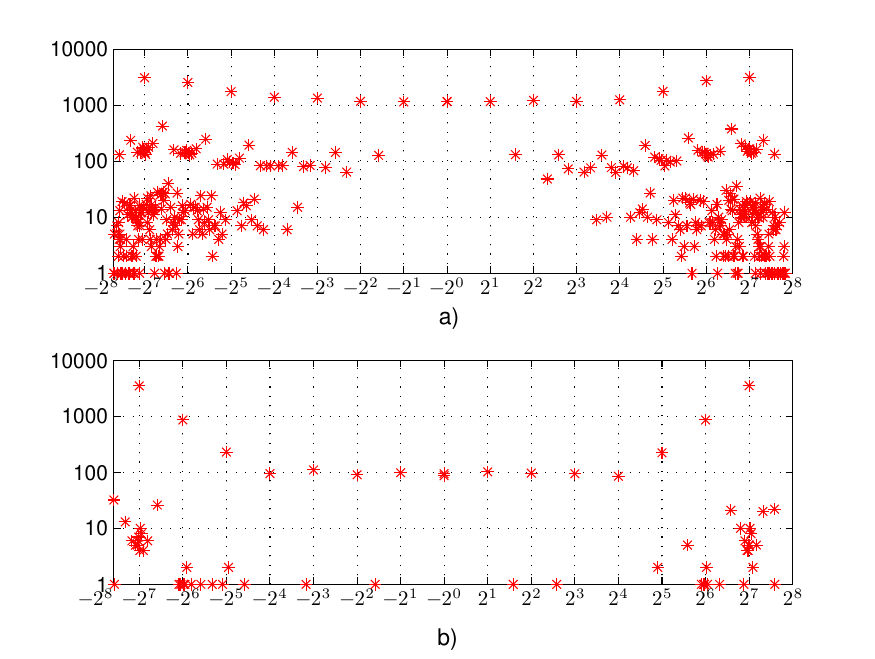}
\end{minipage}
\caption{Distribution of non-zero difference between two recovered
plain-images and the original plain-image: a) the one shown in
Fig.~\ref{figure:KnownPlainTextAttack}c); b) the one shown in
Fig.~\ref{figure:KnownPlainTextAttack}d), where x-axis denotes specific difference
and y-axis denotes the corresponding counts.}
\label{figure:distri_diff}
\end{figure}

\begin{figure}[!htb]
\centering
\begin{minipage}[t]{\figwidth}
\centering
\includegraphics[width=\figwidth]{peppers_e_d25}
a)
\end{minipage}
\begin{minipage}[t]{\figwidth}
\centering
\includegraphics[width=\figwidth]{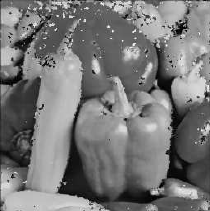}
b)
\end{minipage}
\caption{The enhanced results of applying a $3\times 3$ median filter to the two recovered plain-images: a) the image shown in
Fig.~\ref{figure:KnownPlainTextAttack}c); b) the image shown in
Fig.~\ref{figure:KnownPlainTextAttack}b).} \label{figure:enhanced}
\end{figure}

Distribution of difference between the recovered plain-image, shown
in Fig.~\ref{figure:KnownPlainTextAttack}c), and the corresponding
original plain-image is shown in Fig.~\ref{figure:distri_diff}a).
The similar data on the recovered plain-image shown in
Fig.~\ref{figure:KnownPlainTextAttack}d) is also illustrated in
Fig.~\ref{figure:distri_diff}b) for comparison. With some noise
reduction schemes, the recovered plain-images can be enhanced
further. The results of two images shown in
Figs.~\ref{figure:KnownPlainTextAttack}c) and d) with a $3\times 3$
median filter are shown in Figs.~\ref{figure:enhanced}a) and b),
respectively.

\begin{figure}
\centering
\begin{minipage}[t]{1.5\imagewidth}
\centering
\includegraphics[width=1.5\imagewidth]{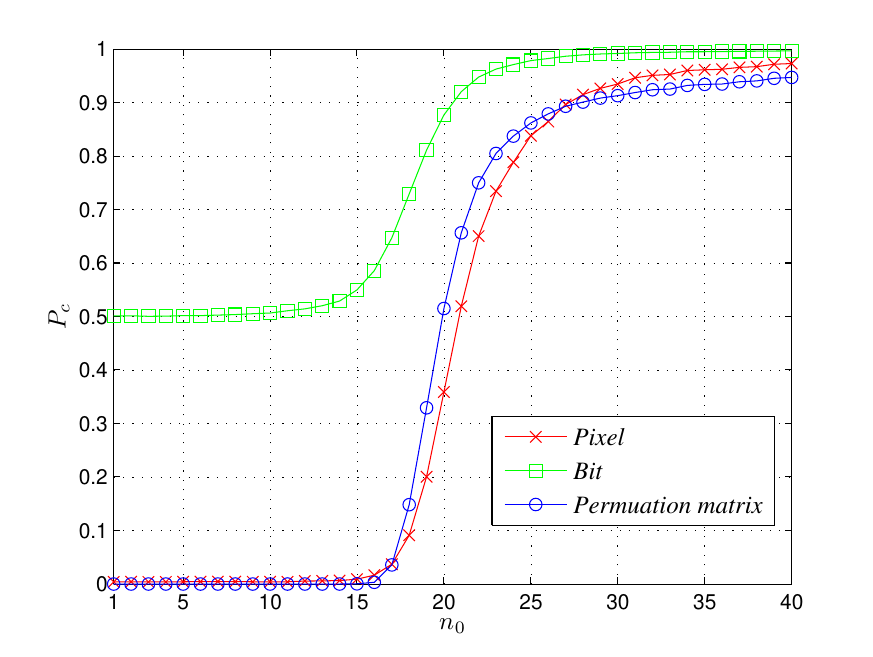}
\end{minipage}
\caption{The percentage of correctly-recovered elements with respect
to the number of known plain-images.}
\label{figure:CorrectPercentage}
\end{figure}

Figure~\ref{figure:CorrectPercentage} shows the percentage of
correctly-recovered elements, including plain-bit, plain-pixel and
the elements of permutation matrix, with respect to the number of
known plain-images. One can see that the breaking performance is
good when $n_0\ge 20$. It can also be observed that the slopes of the
three lines shown in Fig.~\ref{figure:CorrectPercentage} are very
flat when $n_0\ge 25$. This is due to the negative impact incurred
by the strong redundancy in natural images such as the MSBs of
neighboring pixels are the same with a high probability. This point
has been proved quantitatively in \cite{Li:AttackingPOMC2008}. Now,
one can see that the non-uniform distribution of most natural images
has two opposite influences on the final breaking. The experiments
have shown that the above analysis result obtained under the assumption
of uniform distribution also holds for the case of ordinary natural
images.

Obviously, the spatial complexity and the computational
complexity of the proposed known-plaintext attack are only $O(32\cdot MN)$ and
$O(16\cdot n_0\cdot MN)$, respectively.

\subsection{Chosen-plaintext attack}

The chosen-plaintext attack is an attack model for cryptanalysis
which assumes that the attacker has the capability to arbitrarily
choose some plaintexts to be encrypted and observe the corresponding
ciphertexts. The goal of the attack is to optimize the attack
performance considering the special properties of the encryption scheme
when the plaintexts are with a specific structure.

As for the image permutation algorithm under study, at least
$n_c=\lceil \log_2(8MN)\rceil=3+\lceil \log_2(MN)\rceil$ chosen
plain-images are needed to make cardinality of the set in every leaf
node of the binary tree for breaking is equal to one, namely, every
element of the permutation matrix can be recovered exactly. First,
construct an $M\times (8N)$ matrix in domain $\mathbb{Z}_{2^{n_c}}$,
$\bm{B}^{+}=[B^{+}(i,l)]_{i=0, l=0}^{M-1, 8N-1}$, whose elements are
different from each other. Then, the chosen plain-images
$\{\bm{I}_{t}\}_{i=0}^{n_c-1}$ can be chosen/constructed as follows:
$$
I_{t}(i, j)=\sum\nolimits_{k=0}^{7}B^{+}_{t}(i, l)\cdot 2^k,
$$
where $\sum_{t=0}^{n_c-1}B^{+}_{t}(i,l)\cdot 2^t=B^{+}(i, l)$ and
$l=8\cdot j+k$.

\subsection{Some remarks on the performance on the image permutation algorithm under study}

\begin{itemize}
\item \textit{Weak randomness of vectors $\bm{T_M}$ and $\bm{T_N}$}

Obviously, randomness of vectors $\bm{T_M}$ and $\bm{T_N}$ depends
on randomness of the sequences generated by iterating Logistic map.
Note that convincing argument for good randomness of a sequence is
not the so-called complex theories based but whether it can pass a
series of objective tests \cite{Rukhin:TestPRNG:NIST}. As shown in
\cite[Sec. 3.1]{Li:AttackingBitshiftXOR2007}, Logistic maps cannot
serve as a good random number generator. This point can also be
guessed by observing distribution of the trajectory of Logistic map,
which is mainly determined by the control parameter \cite{Kaiwang:PLASymbolic2009}.
For illustration, distribution of two trajectories of Logistic map with
the same initial condition and control parameter used in \cite[Sec.
2]{Ye:Scramble:PRL10} is shown in
Fig.~\ref{fig:distribute4logistic}.

\begin{figure}[!htb]
\centering
\begin{minipage}[t]{\imagewidth}
\centering
\includegraphics[width=\imagewidth]{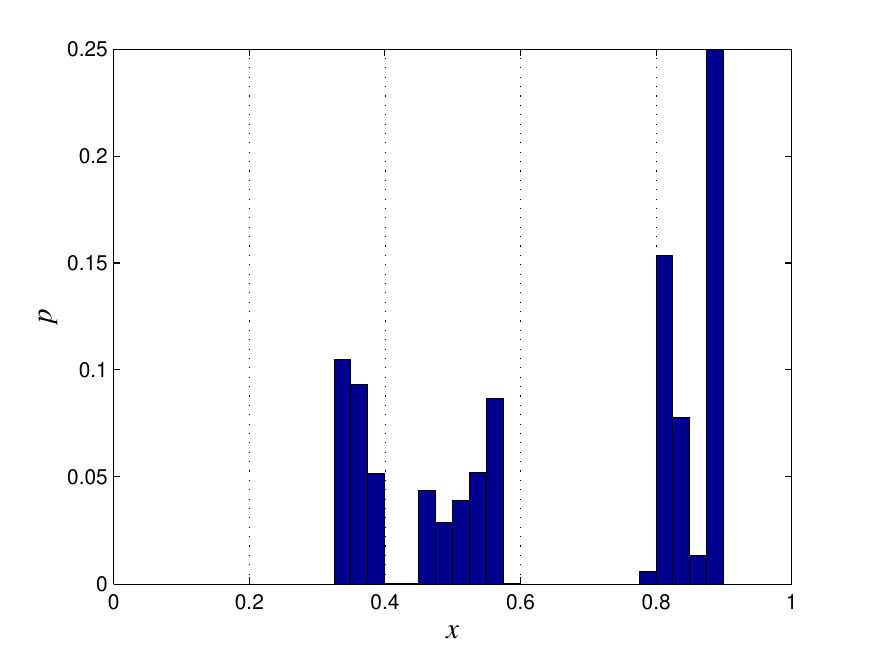}
a)
\end{minipage}
\begin{minipage}[t]{\imagewidth}
\centering
\includegraphics[width=\imagewidth]{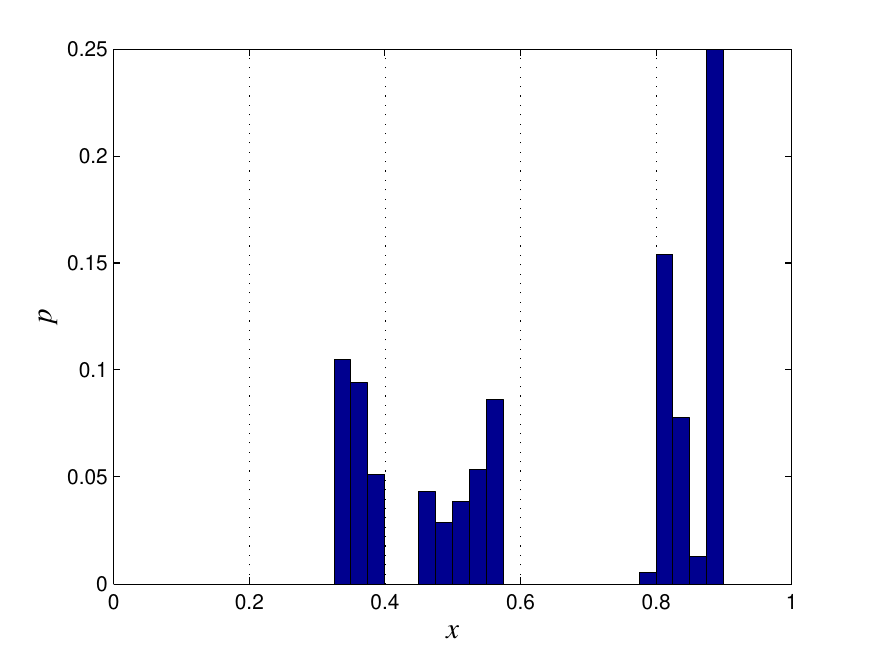}
b)
\end{minipage}
\caption{Distribution of two trajectories of the map
Eq.~(\ref{eq:Logistic}) with control parameter $\mu=3.5786$: a)
$x_0=0.3333$; b) $x_0=0.5656$.} \label{fig:distribute4logistic}
\end{figure}

\item \textit{Fail to encrypt images of fixed value zero or 255}

For both the two images, all elements of the corresponding intermediate matrix
$\bm{B}^*$ are the same, which make all the encryption procedures do nothing.

\item \textit{Insensitivity with respect to changes of plaintext}

Sensitivity with respect to changes of plaintext is very important
for any image encryption scheme since an image and its watermarked
version may be encrypted at the same time. In cryptography, the most
ideal state about the sensitivity is that the change of any single
bit of plaintext will make every bit of the corresponding ciphertext
change with a probability of one half. Unfortunately, the image
permutation algorithm under study does not consider this property at
all.

\item \textit{Equivalent sub-key}

Since $f(x)=f(1-x)$, $x_0$ and $(1-x_0)$ are equivalent sub-keys for
decryption.

\item \textit{Unchanged histogram on plain-bit}

Although the histogram on plain-byte is changed by the image
permutation algorithm under study, the one on plain-bit keep
unchanged, which make the cipher-image still reveal some information
of the plain-image.

\item \textit{Low efficiency}

The generation method of permutation matrixes $\bm{T_M}$ and
$\bm{T_N}$ make the computational complexity of the image
permutation algorithm under study is $O(M\cdot(8N)^2)$, which is
even much larger than that of DES.
\end{itemize}

\section{Known/chosen-plaintext attack on any permutation-only multimedia encryption schemes}
\label{sec:AttackAnyPermuation}

As shown in \cite{Li:AttackingPOMC2008}, no matter what the
permuted elements of multimedia data are, all permutation-only
multimedia encryption schemes on a plaintext of size $M\times N$ can
be represented as
\begin{equation*}
I^*(w(i,j))=I(i,j),
\end{equation*}
where the permutation matrix
$\bm{W}=\left[w(i,j)=(i',j')\in\mathbb{M}\times\mathbb{N}\right]_{M\times
N}$, $\mathbb{M}=\{0,\cdots,M-1\}$ and
$\mathbb{N}=\{0,\cdots,N-1\}$. Then it was analyzed that
permutation-only multimedia encryption can be broken with only
$O(\lceil\log_L(MN)\rceil)$ known/chosen plaintexts, where $L$ is the number of
different elements in the plaintext. The breaking method proposed in
\cite[Sec. 3.1]{Li:AttackingPOMC2008} includes the following three
key steps:

\begin{itemize}
\item \textit{Step 1}: obtain a set containing all possible secret positions for each entry of the
plaintext by comparing every pair of plaintext and the corresponding
ciphertext;

\item \textit{Step 2}: solve the intersection of the different sets
corresponding to each entry of plaintext;

\item \textit{Step 3}: get an estimated version of the permutation
matrix by choosing a secret position of an entry plaintext from the
final set corresponding to it.
\end{itemize}

The operations in \textit{Step 2} make the computational complexity
of the above attack become $O(n_0\cdot MN^2)$. It is found that the
complex intersection operations can be avoided by extending the idea
proposed in Sec.~\ref{sec:KnownPlaintextAttack}, namely constructing
a multi-branch tree, where every node includes $2^{L}+3$ components:
$2^{L}$ pointers holding addresses of $2^{L}$ possible child nodes;
two sets containing some entry positions in plain-image and
cipher-image, respectively; and cardinality of one of the two sets. Let
$\mathbb{B}$, $\mathbb{B}'$, and $|\mathbb{B}|$ denote the last
three items in the root node. Initially, set
$\mathbb{B}=\mathbb{B}'=\mathbb{M}\times \mathbb{N}$, and
$|\mathbb{B}|=MN$. Then, the multi-branch tree can be constructed as
follows.

\begin{itemize}
\item $\forall$ $(i,j)\in \mathbb{B}$, add $(i, j)$ to the first set
of the child nodes to which the $I(i,j)$-th item of the current node
points;

\item $\forall$ $(i,j)\in \mathbb{B}'$, add $(i, j)$ to the second set
of the child nodes to which the $I'(i,j)$-th item of the current node
points.

\item Delete the elements of the sets in the root node and set the
last item of the node as zero.
\end{itemize}

With more pairs of known/chosen plain-images and the corresponding
cipher-images, the multi branch can be updated and expanded
iteratively with the following steps:
\begin{itemize}
\item Search for all nodes whose last item is greater than one;

\item Expand and update each found node with the similar operations shown above.
\end{itemize}

Once construction of the multi-branch tree is completed, the
permutation matrix $\bm{W}$ can be estimated by simply mapping the elements
in the two sets of every leaf node in order.

Finally, combine the performance analysis of the known/chosen-plaintext attack presented in
\cite{Li:AttackingPOMC2008}, we can conclude that any permutation-only multimedia encryption
schemes exerting on plaintext of size $M\times N$ can be efficiently broken with $O(\lceil\log_L(MN)\rceil)$
known/chosen-plaintexts, and spatial complexity and computational
complexity of the attack are $O(MN)$ and $O(n_0\cdot MN)$, respectively. Substitute $n_0$ with $\lceil\log_L(MN)\rceil$,
the complexity becomes $O(\lceil\log_L(MN)\rceil\cdot MN)$, which is lower much than $O(\lceil\log_L(MN)\rceil\cdot MN^2)$, the whole
attack complexity estimated in \cite{Li:AttackingPOMC2008}.

\section{Conclusion}

In this paper, the security of an image permutation encryption
algorithm is analyzed. An optimal method, in term of spatial
complexity and computational complexity, is proposed to break the
binary image permutation algorithm. Furthermore, the method can be
extended to break any permutation-only multimedia encryption schemes
with an optimal performance also. In addition, some specific remarks
on the performance of the image scrambling encryption algorithm
under study are provided. Again, this cryptanalysis paper proves that
the security of a good image encryption scheme should rely on a
combination of traditional cryptography and special properties of
image data, like the schemes proposed in
\cite{Cheng:PartialImageEncryption:IEEETSP2000,Chang:VQImageEncryption:JSS2001}.

\section*{Acknowledgement}

The work of Chengqing Li was partially supported by The Hong Kong Polytechnic
University's Postdoctoral Fellowships Scheme under grant no. G-YX2L.

\bibliographystyle{elsarticle-num}
%\nocite*
\bibliography{SP}
\end{document}